\documentclass[aps,prd,letterpaper,superscriptaddress,nofootinbib]{revtex4-1}
\usepackage{latexsym}
\usepackage{graphicx}
\usepackage{color}
\usepackage{epsfig}
  \newcommand{\be}{\begin{equation}}
  \newcommand{\ee}{\end{equation}}
  \newcommand{\bea}{\begin{eqnarray}}
  
  \newcommand{\eea}{\end{eqnarray}}
  \newcommand{\nn}{\nonumber}
  
  \newcommand{\spur}[1]{\not\! #1 \,}
  \newcommand{\iint}{\int \hspace{-.75em} \int}
  \newcommand{\iiint}{\int \hspace{-.75em} \int \hspace{-.75em} \int}
  \newcommand{\iiiint}{\int \hspace{-.75em} \int \hspace{-.75em} \int \hspace{-.75em} \int}

\begin{document}
\title{On aspects of self-consistency in the Dyson--Schwinger approach to QED and $\lambda (\phi^\star\phi)^2$ theories}
\author{Roberto \surname{Casalbuoni}}
\affiliation{Department of Physics, University of Florence \& INFN-Florence \& Galileo Galilei Institute, Florence, Italy}
\author{Massimo \surname{Ladisa}}
\affiliation{Institut N\'eel, CNRS and UJF, Grenoble, France}
\affiliation{European Theoretical Spectroscopy Facility (ETSF)}
\affiliation{Istituto di Cristallografia, CNR, Bari, Italy}
\author{Valerio \surname{Olevano}}
\affiliation{Institut N\'eel, CNRS and UJF, Grenoble, France}
\affiliation{European Theoretical Spectroscopy Facility (ETSF)}
\affiliation{Istituto di Cristallografia, CNR, Bari, Italy}

\begin{abstract}
We investigate some aspects of the self-consistency in the Dyson--Schwinger 
approach to both the QED and the self-interacting scalar field theories. 
We prove that the set of the Dyson--Schwinger equations, together with 
the Green--Ward--Takahashi identity, is equivalent to the analogous set 
of integral equations studied in condensed matter, namely many-body 
perturbation theory, where it is solved self-consistently and 
iteratively. In this framework, we compute the non-perturbative solution 
of the gap equation for the self-interacting scalar field theory.
\end{abstract}
\maketitle

\section{Introduction}

Nonperturbative approaches to quantum field theory (QFT) 
allow for a better understanding of some general properties 
of the exact scattering amplitudes. They historically pursued after 
the formulation of quantum electrodynamics (QED) to analyse the 
asymptotic behaviour of renormalized operators at short distance 
(see, for instance, Ref.~\cite{GellMann:PR95}).
  
Beyond that, they give considerable informations in detailed 
studies of the structure of higher order approximations in theories 
where the perturbation expansion, along with a strong coupling, 
fails in analysing the short distance behaviour of the relevant 
theory operators.
This approach (referred 
to as Dyson--Schwinger) has been historically introduced 
by a number of authors
\cite{Dyson:PR75_1,Dyson:PR75_2,Schwinger:PNAS37_1,Schwinger:PNAS37_2,Feynman:PR84}.
 
Since then, it has inspired plenty of papers on the subject 
(see, for instance, Refs.~\cite{Curtis:1990zs,Curtis:1993py,Dong:1994jr,Roberts:PPNP33,Munczek:PRD52,Bender:2002as,Zwanziger:2002ia,Bhagwat:2004hn,Binosi:JHEP}). 
Among them we quote those dealing with non-abelian theories 
\cite{Roberts:PPNP33,Munczek:PRD52,Bender:2002as,Zwanziger:2002ia,Bhagwat:2004hn,Binosi:JHEP}, 
such as quantum chromodynamics (QCD), where both the 
aymptotic freedom (large coupling values at low energy) 
and the confinement of the Lagrangian fields within the 
asymptotic ones intrinsically require a phenomenological 
analysis beyond the naive perturbative expansion. 
In these processes the short-distance contributions can be 
computed, to some extent, by using the factorization 
approximation for the local operators in the effective 
Hamiltonian, as argued by Bjorken on the basis of 
color-transparency \cite{Bjorken:NPB11}.

Large coupling and field confinement also characterize 
solid state physics phenomenology, where a screened 
many-body interaction occurs at about the Fermi energy among 
a large number of electrons embedded in the crystal lattice. 
In condensed matter singularities appear in the 
perturbation series of the correlation energy of a fully 
degenerate Fermi--Dirac system with Coulomb interaction, 
otherwise named homogeneous electron gas (HEG): 
in Ref.~\cite{GellMann:PR106} the 
large divergent logarithms are resummed according to the 
general scheme introduced by Feynman in QFT 
\cite{Feynman:PR84}. In spite of its simplicity, 
HEG is fundamental for computing the correlation energy 
of a wide class of complex systems within the so called 
local-density approximation \cite{Aryasetiawan:RPP61}. 
 
Also within this context, a Green's function 
theory approach, called (improperly) many-body perturbation 
theory (MBPT), based on a formalism of second quantization of 
operators, has been considered \cite{Baym:PR124,Baym:PR127}. 
The fundamental degree of freedom is the Green's function or 
propagator, which represents the probability amplitude for 
the propagation of an electron. As in any other QFT, the 
many-body system can be expanded in perturbation theory, 
with the coupling being the many-body interaction term. 
The Green's function (as well as any other quantity of the 
theory, such as the self-energy or the polarization) can be 
calculated at a given order of perturbation theory. A Feynman 
diagrammatic analysis is hence possible. The theory at the first 
order is equivalent to the Hartree--Fock theory. However the 
coupling is not small (compared, for example, to the 
electron-ion interaction) and the expansion does not converge. 
The second order is not necessarily smaller than the first. 
Hence one needs to resort to more complicated methods to 
solve the theory. Beyond partial resummations of diagrams 
to all orders, iterative methods have been preferred.
 
Historically a Dyson--Schwinger approach 
has been introduced to account for such a non-perturbative phenomenology 
and to compute optical and electronic 
properties of complex systems by means of iterative schemes. 
The alternative formulation 
in terms of functional derivatives reduces the many-body problem 
to the solution of a coupled set of nonlinear integral equations, 
whose characteristic feature is that, besides the single-particle 
Green's function, a whole hierarchy of equations involving 
higher order Green's functions is generated. A truncation 
of Feynman diagrammatics corresponds to the custom of 
replacing this hierarchy of equations by another coupled set of 
nonlinear equations connecting the single-particle Green's 
function to the mass, polarization and vertex operators, 
often referred to as the Hedin's equations in condensed matter 
\cite{Hedin:PR139}, to be solved iteratively until self-consistency 
is achieved\footnote{The approach described here is akin to 
the {\it truncation} procedure of the infinite tower of Dyson--Schwinger 
equations in particle physics (see, for instance, Ref. \cite{Curtis:1990zs}).}. 
Here the higher order Green's functions dependence of the relevant 
quantities is recasted within the functional derivative of the 
mass operator with respect to the fermion single-particle Green's 
function. The latter quantity is proved to equate the Bethe--Salpeter 
kernel for the two-body into two-body rescattering. 
Both the diagrammatic truncation and the nonlinear coupled set of 
equations correspond to a nonperturbative approximation to the 
solution of the many-body problem. 
 
So far, nobody has solved the Hedin's equations for a real system, 
since the nonlinearity of the equation involving the 
Bethe--Salpeter kernel is computationally demanding. Approximations 
are required to simplify the problem. Among the 
most widely used computation schemes it is worth mentioning the 
so called {\it GW approximation}\footnote{The name {\it GW} stands 
for the {\it product} of the {\it G} Green's function and the 
{\it W} dressed Coulomb interaction while computing the mass operator 
$\Gamma G W$, being $\Gamma$ the vertex operator as it appears in the 
Lagrangian ($\Gamma \approx \bf{I}$ when magnetic and relativistic 
effects are neglected).}, 
where the vertex operator is simplified in the self-energy 
evaluation at the beginning, and the Bethe--Salpeter equation accounts for the vertex 
corrections within this approach (see, for instance, the reviews in 
\cite{Aryasetiawan:RPP61,Strinati:RNC11}). 
  
The analogies between these two 
fields, spanning ultrahigh and ultralow energies, have 
been cross fertilizing methodologies and approaches. Along 
with that, this paper aims at spotting: i) the equivalence 
between the Hedin's equations set and the Dyson--Schwinger 
approach to QFT, where the role played by the underlying 
gauge symmetry is crucial in relating the Bethe--Salpeter 
kernel to the functional derivative of the mass operator 
with respect to the fermion Green's function (throughout the 
Green--Ward--Takahashi identity 
\cite{Ward:1950xp,Green:1953te,Takahashi:1957xn}); 
ii) the iterative scheme solution of the Dyson--Schwinger 
equations, in the spirit of the condensed matter methods, 
by means of a S-matrix unitarity inspired ansatz on the 
mass operator, accounting for the nonlinearity of the present 
approach. 
Both points should be regarded as aspects of the 
self-consistency of the Dyson--Schwinger approach to the 
many-body quantum field theory, with regard to the actual 
theory investigated, in the context of the aforementioned 
S-matrix unitarity inspired picture. Indeed, in QFT the 
S-matrix satisfies the unitarity condition, actually more 
fundamental than the concept of Hamiltonian and wave 
functions (see, for instance, 
Refs.~\cite{GellMann:PR95,Landau:1956zr,Wilson:PR179,DellAntonio:1971zt} 
and references therein). 
 
In that respect, we point out two 
examples (QED and the self-interacting scalar field 
theory), different on the physical basis but akin as 
to the solution scheme. The QED example is intended to 
formally bridge the condensed matter scenario to its 
underlying fundamental theory: throughout the paper, 
a QFT analysis to the topics has been preferred to the 
functional one, in order to spot this idea. A functional 
approach is, of course, viable. On the other hand, the 
gap equation arising in the self-interacting scalar field 
theory is probably the simplest example to show how to 
implement the iterative scheme to the numerical solution of 
the Dyson--Schwinger equations.
 
The plan of the paper is as follows. In the next section we
discuss the general formalism of Dyson--Schwinger approach in QED. 
In section 3, the Green--Ward--Takahashi identity is proved 
to be equivalent to the definition of the Bethe--Salpeter kernel as 
the functional derivative of the mass operator with respect to the 
Green's function. Section 4 is devoted to
the calculation of the Dyson--Schwinger equations for the 
self-interacting scalar field theory. Finally in section 5 
we compute the non-perturbative solution to the gap equation 
arising for the latter case and draw our conclusions.

\section{The set of Dyson--Schwinger equations in QED}
It is worth to recall the Dyson--Schwinger equations for QED. 
Hereafter, latin letters shall refer to noninteracting quantities, while 
calligraphic symbols represent interacting ones. Thus, $\psi$ represents the free 
fermionic field, while $\Psi$ is the exact fermionic field; analogously for the 
photonic field $A$ (${\mathcal A}$). Accordingly, $G$ (${\mathcal G}$) is the free 
(exact) fermionic Green's function, while 
$D$ (${\mathcal D}$) corresponds to the free (exact) photon propagator. The 
electromagnetic current $j$ is defined according to Ref.~\cite{BerestetskiiQED}, 
{\it i.e.} ${\displaystyle j^\mu(x) \stackrel{def.}{=} \bar\Psi(x)\gamma^\mu\Psi(x)}$. 
They read: 
\bea
\label{eqn:G_def}
G^\alpha_\beta(x,y) &\stackrel{def.}{=}& 
-i \langle 0 | T\left( \psi^\alpha(x) \overline \psi_\beta(y) \right) | 0 \rangle \;\; , \hspace*{1cm} 
\left( i \spur \partial_x - m~{\mathcal I} \right) \psi(x) ~=~ 0 \;\; , \\
\label{eqn:D_def}
D^{\mu\nu}(x,y) &\stackrel{def.}{=}& 
+i \langle 0 | T\left( A^\mu(x) A^{\dagger \nu}(y) \right) | 0 \rangle \;\; , \hspace*{1cm} 
\left( \partial^\mu \partial^\nu - \partial_\alpha \partial^\alpha g^{\mu\nu} \right) A_\nu(x) =  0 \;\; , \\
\label{eqn:GG_def}
{\mathcal G}^\alpha_\beta(x,y) &\stackrel{def.}{=}& 
-i \langle 0 | T\left( \Psi^\alpha(x) \overline \Psi_\beta(y) \right) | 0 \rangle \;\; , \hspace*{1cm} 
\left( i \spur \partial_x - m~{\mathcal I} \right) \Psi(x) ~=~ e~ {\mathcal {\not\! \hspace{-.25em} A \,}}(x) \Psi(x) \;\; , \\
\label{eqn:DD_def}
{\mathcal D}^{\mu\nu}(x,y) &\stackrel{def.}{=}& 
+i \langle 0 | T\left( {\mathcal A}^\mu(x) {\mathcal A}^{\dagger \nu}(y) \right) | 0 \rangle \;\; , \hspace*{1cm} 
\left( \partial^\mu \partial^\nu - \partial_\alpha \partial^\alpha g^{\mu\nu} \right) {\mathcal A}_\nu(x) = -4\pi e~j^\mu(x) \;\; , \\
\label{eqn:Gamma_def}
\langle 0 | T\left( {\mathcal A}^\mu(x) \Psi^\alpha(y) \overline\Psi_\beta(z) \right) | 0 \rangle 
&\stackrel{def.}{=}& e~\iiint d^4x^\prime d^4y^\prime d^4z^\prime \left\{ 
{\mathcal G}^{\beta^\prime}_\beta(z,z^\prime)~\Gamma^{\mu^\prime \alpha^\prime}_{\beta^\prime}(z^\prime,y^\prime,x^\prime)~{\mathcal G}^\alpha_{\alpha^\prime}(y^\prime,y)~{\mathcal D}^\mu_{\mu^\prime}(x^\prime,x) \right\} \;\; , \\
\label{eqn:M_P_defs}
{\mathcal M}^\alpha_\beta(x,y) &\stackrel{def.}{=}& G^{-1\alpha}_\beta(x,y) - {\mathcal G}^{-1\alpha}_\beta(x,y) \;\; , 
\hspace*{1cm}  {\mathcal P}^{\mu\nu}(x,y) \stackrel{def.}{=} D^{-1\mu\nu}(x,y) - {\mathcal D}^{-1\mu\nu}(x,y) \;\; ,
\eea
where a sum over repeated indices is understood.
The relevant equation for ${\mathcal M}$ operator is obtained by applying $G^{-1~\alpha}_{~~\beta}(x,y)$ 
$\left(\displaystyle{ \stackrel{def.}{=} \delta^4(x-y) \left( i\spur \partial_y - m {\mathcal I} \right)}^\alpha_\beta \right)$ 
to ${\mathcal G}$ operator defined in Eq.~(\ref{eqn:GG_def}) and by using Eq.~(\ref{eqn:Gamma_def}) 
for the vertex $\Gamma$ \footnote{Details on this calculation are carefully reported in Ref.~\cite{BerestetskiiQED}.}. It reads:
\be
\label{eqn:M_eq}
{\mathcal M}^\alpha_\beta(x,y) = -i e^2 \iint d^4x^\prime d^4y^\prime ~\gamma^\alpha_{\alpha^\prime\mu} 
~{\mathcal G}^{\alpha^\prime}_{\beta^\prime}(x,x^\prime) 
~\Gamma^{\beta^\prime}_{\beta\nu}(x^\prime,y,y^\prime) 
~{\mathcal D}^{\nu\mu}(y^\prime,x) \; ;
\ee
analogously for ${\mathcal P}$ operator one gets:
\be
\label{eqn:P_eq}
{\mathcal P}^{\mu\nu}(x,y) = 4 \pi i e^2 \iint d^4x^\prime d^4y^\prime ~\gamma^{\alpha\mu}_{\alpha^\prime} 
~{\mathcal G}^{\alpha^\prime}_{\beta^\prime}(x,x^\prime) 
~\Gamma^{\beta^\prime\nu}_{\delta^\prime}(x^\prime,y^\prime,y)
~{\mathcal G}^{\delta^\prime}_{\alpha}(y^\prime,x) \; .
\ee
 
Beside these, another equation is needed for the vertex $\Gamma$:
\bea
\label{eqn:Gamma_eq}
i \Gamma^{\alpha\mu}_\beta(x,y,z) &\stackrel{def.}{=}& i \gamma^{\alpha\mu}_\beta\delta^4(x-z)\delta^4(y-z) + \nn \\
&+& i \iiiint d^4x^\prime d^4y^\prime d^4x^{\prime\prime} d^4y^{\prime\prime} 
~{\mathcal G}_{\alpha^\prime}^{\alpha^{\prime\prime}}(x^{\prime\prime},x^\prime) 
~\Gamma^{\alpha^\prime\mu}_{\beta^\prime}(x^\prime,y^\prime,z) 
~{\mathcal G}^{\beta^\prime}_{\beta^{\prime\prime}}(y^\prime,y^{\prime\prime})
~{\bf K}^{\beta^{\prime\prime}\alpha}_{\alpha^{\prime\prime}\beta}(y,y^{\prime\prime},x,x^{\prime\prime}) \; ,
\eea
where ${\bf K}$, the Bethe--Salpeter kernel, accounts for all the possible 
contributions coming from the fermion-fermion rescattering, except for those already 
embodied within the exact propagators ${\mathcal G}, {\mathcal D}$ and  
vertex $\Gamma$\footnote{In a functional approach they are named irreducible diagrams.}.
  
Equations (\ref{eqn:M_P_defs},\ref{eqn:M_eq},\ref{eqn:P_eq},\ref{eqn:Gamma_eq}) 
are also reported in a diagrammatic representation:

\begin{figure}[ht!]
\begin{center}
\epsfig{file=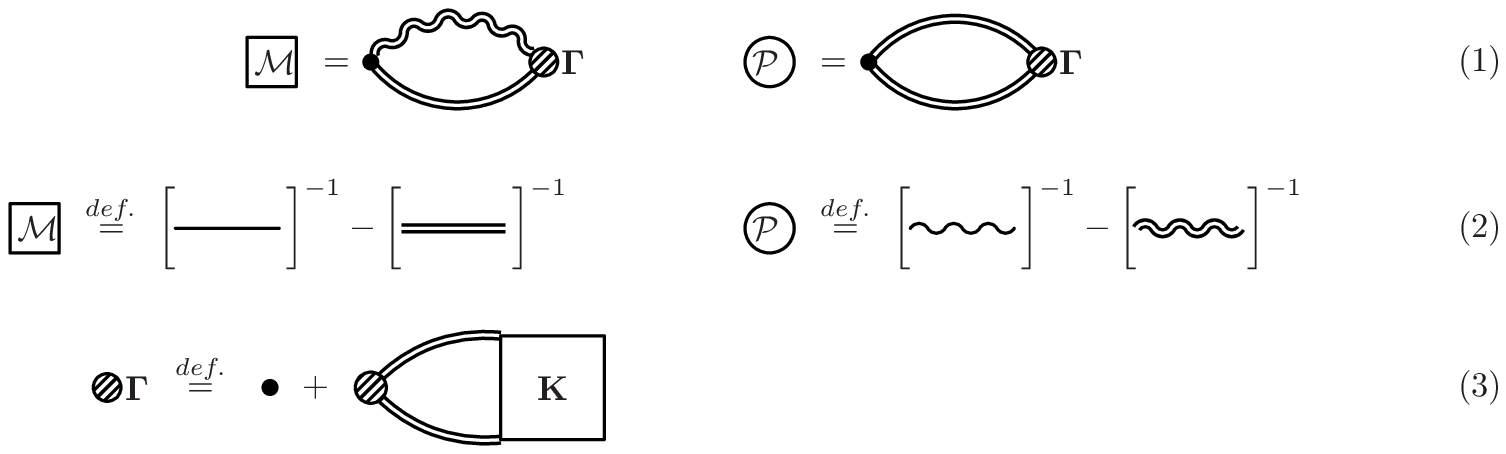,height=5cm,width=15cm}
\end{center}
\end{figure}
 
As a matter of fact, the set of five integral equations depicted above 
depends on six unknowns, {\it i.e.} 
${\mathcal G}, {\mathcal D}, {\mathcal P}, {\mathcal M}, \Gamma$ and ${\bf K}$. 
Thus, an additional equation is needed to close the system and to approach 
a solution. Two schemes have been assessed to accomplish with such an issue.
In condensed matter the Bethe--Salpeter kernel is written as the functional 
derivative of the mass operator with respect to the Green's function: 
${\displaystyle {\bf K} = \delta {\mathcal M} / \delta {\mathcal G}}$ 
\cite{Aryasetiawan:RPP61,Hedin:PR139,Strinati:RNC11}. In QFT, 
the same definition formally holds. Indeed, in the framework of 
Cornwall-Jackiw-Tomboulis approach \cite{Cornwall:1974vz}, it is known 
that the second functional derivative of the action with respect to the 
(bilocal) field equals the Bethe--Salpeter kernel \cite{McKay:1989rk}. 
On the other hand, in Ref.~\cite{Munczek:PRD52} such a second functional 
derivative is proved to be the functional derivative of the mass 
operator with respect to the fermion Green's function. While such a 
relation is crucial for the self-consistency of the coupled set of 
nonlinear equations, in order to avoid a hierarchy of higher 
order Green's functions, a functional derivative is computationally 
demanding and a diagrammatic, although non-perturbative, expansion of 
the Bethe--Salpeter kernel is usually preferred\footnote{In particle physics, 
this approach is usually referred to as {\it truncation} of Dyson--Schwinger 
equations.} 
\cite{Curtis:1990zs,Curtis:1993py,Dong:1994jr,Roberts:PPNP33,Bender:2002as,Zwanziger:2002ia,Bhagwat:2004hn,Binosi:JHEP}.

\section{The Green--Ward--Takahashi identity.}
 
Under a small phase (gauge) shift on the $\Psi$ operators, $\delta\chi$, the 
${\mathcal G}(x,x^\prime)$ operator is shifted by 
${\displaystyle i e {\mathcal G}(x,x^\prime) \left[\delta\chi(x) - \delta\chi(x^\prime)\right]}$, 
according to the definition of Eq.~(\ref{eqn:GG_def}). On the other hand, 
$\delta{\mathcal G}(x,x^\prime)$ can be directly computed as the coupling of the 
(small) gauge field $-\partial_\mu\delta\chi$ to the fermionic current\footnote{Polarization does 
not affect the gauge field, since the former is a transverse tensor while the latter is longitudinal. 
Once again, details can be found in Ref.~\cite{BerestetskiiQED}.}. By equating the former and 
the latter quantities, one gets the Green's equation \cite{Green:1953te} 
either in its integral formulation, {\it i.e.}
\be
\label{eqn:GWT_int}
i e {\mathcal G}^\alpha_\beta(x,x^\prime) \left[\delta\chi(x) - \delta\chi(x^\prime)\right] = 
e \iiint d^4y d^4y^\prime d^4z~ {\mathcal G}^\alpha_{\alpha^\prime}(x,y)~ 
\partial_z^\mu \Gamma^{\alpha^\prime}_{\beta^\prime\mu}(y,y^\prime,z)~ 
{\mathcal G}^{\beta^\prime}_\beta(y^\prime,x^\prime)~\delta\chi(z)\;\ ,
\ee
or in its differential one ${\displaystyle -i e {\mathcal G}^{-1}(x,x^\prime) 
\left[\delta\chi(x) - \delta\chi(x^\prime)\right] = e \int d^4z~ \partial_z^\mu 
\Gamma_\mu(x,x^\prime,z)~ \delta\chi(z)}$. We will need both forms in the sequel. The 
Ward--Takahashi identity corresponds to the soft photon limit in momentum space 
\cite{Ward:1950xp,Takahashi:1957xn}.
 
To accomplish with the task of completing the set of Dyson--Schwinger equations, 
we replace ${\mathcal G}^{-1}$ in the differential form of  
Green--Ward--Takahashi equation, according to the mass operator definition of 
Eq.~(\ref{eqn:M_P_defs}), and we notice that the non-interacting propagator $G$ is 
gauge invariant, therefore 
\be
\label{eqn:G_GWTI}
i G^{-1}(x,x^\prime)\left[\delta\chi(x) - \delta\chi(x^\prime)\right] ~=~
\int d^4z~ \delta\chi(z)~ \partial_z^\mu \left[\gamma_\mu \delta^4(x-z) 
\delta^4(x^\prime-z) \right] ~=~ 0\;\; . 
\ee
Thus:
\bea
\label{eqn:M_GG_eq}
&& i e {\mathcal M}(x,x^\prime)\left[\delta\chi(x) - \delta\chi(x^\prime)\right] ~=~ 
e \int d^4z~ \delta\chi(z)~ \partial_z^\mu 
\left[- \gamma_\mu \delta^4(x-z) \delta^4(x^\prime-z) +  \Gamma_\mu(x,x^\prime,z)\right] ~=~\nn \\
&=& e \int d^4z~\delta\chi(z)~
\iiiint d^4\hat x^\prime d^4\hat y^\prime d^4\hat x^{\prime\prime} d^4\hat x^{\prime\prime} 
{\mathcal G}(\hat x^\prime,\hat x^{\prime\prime})~
\partial_z^\mu \Gamma_\mu(\hat x^{\prime\prime},\hat y^{\prime\prime},z)~
{\mathcal G}(\hat y^{\prime\prime},\hat y^\prime)~ {\bf K}(\hat y^\prime,x^\prime,\hat x^\prime,x) ~=~ \nn \\
&=&
\iint d^4\hat x^\prime d^4\hat y^\prime~ \delta{\mathcal G}(\hat x^\prime,\hat y^\prime)~
{\bf K}(\hat y^\prime,x^\prime,\hat x^\prime,x) \;\ ,
\eea
where in the latter equations we used the vertex definition of Eq.~(\ref{eqn:Gamma_eq}) and 
the integral form of Green--Ward--Takahashi relation of Eq.~(\ref{eqn:GWT_int}). It can be 
easily seen that the l.h.s. of Eq.~(\ref{eqn:M_GG_eq}) is the shift of the mass operator 
under a gauge transform\footnote{Indeed under a gauge shift:
${\displaystyle \delta\left\{\int d^4y~{\mathcal G}(x,y) {\mathcal G}^{-1}(y,x^\prime) \right\}~=~0}$. 
By expanding the latter gauge variation and by noticing that the non-interacting propagator $G$ 
(unlike ${\mathcal G}$ and ${\mathcal M}$) is gauge invariant, we achieve the result.},
 {\it i.e.} $\delta{\mathcal M}(x,x^\prime)$. 
  
In conclusion 
\be
\label{eqn:DeltaM_QED}
\delta{\mathcal M}(x,x^\prime)~=~\iint d^4\hat x^\prime d^4\hat y^\prime~ 
\delta{\mathcal G}(\hat x^\prime,\hat y^\prime)~{\bf K}(\hat y^\prime,x^\prime,\hat x^\prime,x)\;\; .
\ee 
The meaning of this equation is twofold. It states that: i) the gauge shift on 
the mass operator is linearly dependent on the gauge shift of the fermion propagator throughout 
the Bethe--Salpeter kernel; ii) for the same reason, the Bethe--Salpeter kernel 
is gauge invariant. Moreover, the equation written above completes the set of Dyson--Schwinger 
equations and makes finding a solution a viable problem. 

\section{The set of Dyson--Schwinger equations in the $\lambda(\phi^\star\phi)^2$ theory.}
The set of Dyson--Schwinger equations for self-interacting scalar field theory is simpler 
than the analogous in QED. Hereafter $\varphi$ ($\phi$) represents the free (exact) bosonic 
field and, accordingly, $D$ (${\mathcal D}$) is the free (exact) boson Green's function. It reads:
\bea
\label{eqn:D4_def}
D(x,y) &\stackrel{def.}{=}& 
+i \langle 0 | T\left( \varphi(x) \varphi^\star(y) \right) | 0 \rangle \;\; , \hspace*{1cm} 
\left(\Box_x + m^2\right) \varphi(x) =  0 \;\; , \\
\label{eqn:DD4_def}
{\mathcal D}(x,y) &\stackrel{def.}{=}& 
+i \langle 0 | T\left( \phi(x) \phi^\star(y) \right) | 0 \rangle \;\; , \hspace*{1cm} 
\left(\Box_x + m^2\right) \phi(x) = \frac{\lambda}{2!}~\phi^\star(x)\phi^2(x) \;\; , \\
\label{eqn:Pi_def}
\langle 0 | T\left( \phi(x) \phi(y) \phi^\star(z) \phi^\star(w) \right) | 0 \rangle 
&\stackrel{def.}{=}& -i \lambda~\times \nn \\
&\times& \iiiint d^4x^\prime d^4y^\prime d^4z^\prime d^4w^\prime \left\{ 
{\mathcal D}(w,w^\prime)~{\mathcal D}(z,z^\prime)~\Pi(w^\prime,z^\prime,y^\prime,x^\prime)~
{\mathcal D}(y^\prime,y)~{\mathcal D}(x^\prime,x) \right\} \;\; , \\
\label{eqn:M_def}
{\mathcal M}(x,y) &\stackrel{def.}{=}& D^{-1}(x,y) - {\mathcal D}^{-1}(x,y) \;\; ,
\eea
being 
${\displaystyle \Box_x \stackrel{def.}{=} \frac{\partial}{\partial x^\mu}\frac{\partial}{\partial x_\mu}}$. 
The latter and the former sets are akin as to the derivation. 
Let us consider, for instance, the mass operator equation. By applying $D^{-1}$ to ${\mathcal D}$ 
(eqs. \ref{eqn:D4_def},\ref{eqn:DD4_def}) one gets:
\be
\label{eqn:Mphi4_0}
\int d^4y~ D^{-1}(x,y)~{\mathcal D}(y,x^\prime)~=~\delta^4(x-x^\prime) + i \frac{\lambda}{2!}~
\langle 0 | T\left( \phi(x)^2 \phi^\star(x) \phi^\star(x^\prime) \right) | 0 \rangle \;\; .
\ee
Finally, by using Eq.~(\ref{eqn:Pi_def}), the mass operator is computed:
\be
\label{eqn:Mphi4_eq}
{\mathcal M}(x,x^\prime)~=~\frac{\lambda^2}{2!}~
\iiint d^4\hat y^\prime d^4\hat y^{\prime\prime} d^4\hat x^{\prime\prime}~
{\mathcal D}(x,\hat y^{\prime\prime})~{\mathcal D}(x,\hat x^{\prime\prime})~
\Pi(\hat y^{\prime\prime},\hat x^{\prime\prime},\hat y^\prime,x^\prime)~
{\mathcal D}(\hat y^\prime,x)\;\ .
\ee
The vertex equation introduces a Bethe--Salpeter kernel for the two-body rescattering:
\bea
\label{eqn:Pi_eq}
\Pi(y^\prime,x^\prime,y,x)&=&\delta^4(y^\prime-x^\prime)\delta^4(y^\prime-y)\delta^4(y^\prime-x) +\nn \\
&+&\iiiint d^4\hat x^\prime d^4\hat x^{\prime\prime} d^4\hat y^\prime d^4\hat y^{\prime\prime}~
\Pi(y^\prime,x^\prime,\hat y^{\prime\prime},\hat x^{\prime\prime})~
{\mathcal D}(\hat y^{\prime\prime},\hat y^\prime)~{\mathcal D}(\hat x^{\prime\prime},\hat x^\prime)~
{\bf K}(\hat y^\prime,\hat x^\prime,y,x) \;\ .
\eea
 
Although the self-interacting scalar field theory is not a gauge theory, the gauge shifts 
of mass operator and propagator (under a phase shift of $\phi$ operator) are related in 
the same fashion of QED case\footnote{For a strictly neutral 
$\phi$ field, {\it i.e.} $\phi^\star=\phi$, an additional (Schwinger) term arises. 
The absence of a global gauge symmetry for this case has been also pointed out in 
\cite{PauliRMP13}.}:
\be
\label{eqn:DeltaM_phi4}
\delta{\mathcal M}(x,x^\prime)~=~\iint d^4\hat x^\prime d^4\hat y^\prime~ 
\delta{\mathcal D}(\hat x^\prime,\hat y^\prime)~{\bf K}(\hat y^\prime,x^\prime,\hat x^\prime,x)\;\; .
\ee

The latter equation completes the set of Dyson--Schwinger ones , Eqs.~(\ref{eqn:M_def}, \ref{eqn:Mphi4_eq}, \ref{eqn:Pi_eq}) for the self-interacting scalar field theory. 
They can be recapitulated in the following diagrammatics where $\delta\left[\cdot\cdot\cdot\right]$ 
means the gauge shift of the quantity within brackets:

\begin{figure}[ht!]
\label{fig:DSE4}
\begin{center}
\epsfig{file=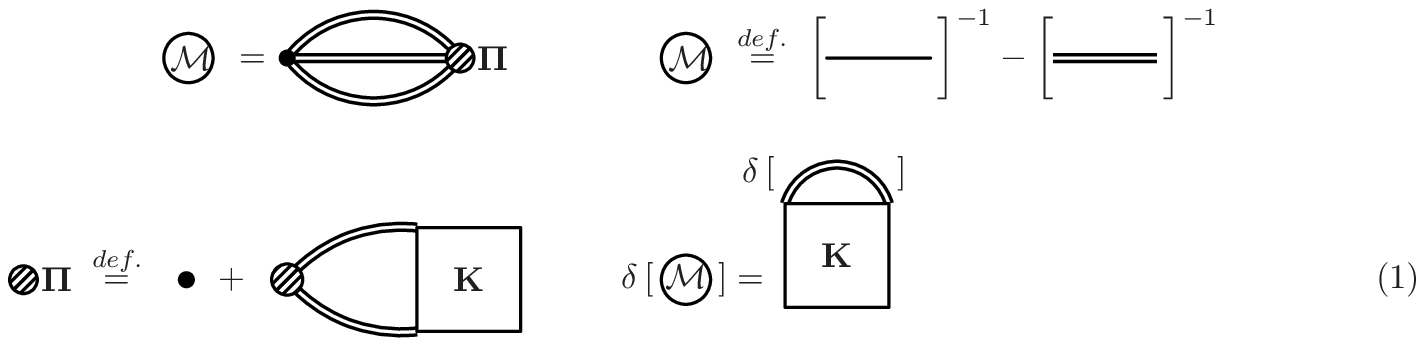,height=4cm,width=16cm}
\end{center}
\end{figure}
\section{Results and discussion}
Hereafter we shall assume that the space-time is homogeneous. Therefore 
the energy-momentum conservation law holds in momentum space, where the 
relevant quantities defined in the previous section simplify accordingly\footnote{\label{foot:DSE4_homo}
For instance, the propagator in the momentum space reads 
$\tilde{\mathcal D}(k,k^\prime)$; it reduces to 
$(2\pi)^4\tilde{\mathcal D}(k)~\delta^4(k-k^\prime)$. On the same footing, the gauge shift 
$\delta\tilde{\mathcal D}(p,p^\prime)=i e \int d^4q/(2\pi)^4~\delta\tilde\chi(q)~
\left[\tilde{\mathcal D}(p+q,p^\prime)-\tilde{\mathcal D}(p,p^\prime-q)\right]$ 
is shortened to $\delta\tilde\chi(p^\prime-p)~
\left[\tilde{\mathcal D}(p^\prime)-\tilde{\mathcal D}(p)\right]$. 
Analogously for the mass operator. The Bethe--Salpeter kernel is also simplified: 
${\displaystyle \tilde{\bf K}(p,q,p^\prime,q^\prime)~=~(2\pi)^4~\delta^4(p+q-p^\prime-q^\prime)~\tilde {\bf K}(p,q,p^\prime,q^\prime)}$.}. 
Thus the set of Dyson--Schwinger equations reduces to:
\bea
\label{eqn:DSE4_homo}
\tilde{\mathcal M}(p) &\stackrel{def.}{=}& \tilde D^{-1}(p)-\tilde{\mathcal D}^{-1}(p) \;\ , \nn \\
\tilde{\mathcal M}(p) &=& \frac{\lambda^2}{2!}\iint\frac{d^4\sigma}{(2\pi)^4}\frac{d^4\eta}{(2\pi)^4}~
\tilde{\mathcal D}(\sigma)~\tilde{\mathcal D}(\eta)~\tilde\Pi(-\sigma,-\eta,p-\sigma-\eta,-p)~
\tilde{\mathcal D}(\sigma+\eta-p)\;\ , \nn \\
\tilde\Pi(p+\sigma,q-\sigma,p,q) &\stackrel{def.}{=}& {\bf I} + \int\frac{d^4 k}{(2\pi)^4}~
\tilde\Pi(p+\sigma,q-\sigma,p+k,q-k)~\tilde{\mathcal D}(q-k)~\tilde{\mathcal D}(p+k)~
\tilde{\bf K}(p+k,q-k,p,q)\;\ , \nn \\
\tilde{\mathcal M}(p+q) - \tilde{\mathcal M}(p) &=& \int\frac{d^4 k}{(2\pi)^4}~
\left[\tilde{\mathcal D}(k) - \tilde{\mathcal D}(k-q)\right]~\tilde{\bf K}(k-q,p+q,k,p)\;\ .
\eea
For the sake of clarity, we prove the last equation in (\ref{eqn:DSE4_homo}). 
We start from the last equation in (\ref{eqn:M_GG_eq}) for the self-interacting scalar 
field theory:
\bea
\label{eqn:DSE4_homo_proof}
\delta{\mathcal M}(x,x^\prime) &=&
\iint d^4\hat x^\prime d^4\hat y^\prime~ 
\delta{\mathcal D}(\hat x^\prime,\hat y^\prime)~{\bf K}(\hat y^\prime,x^\prime,\hat x^\prime,x) \;\; 
\Longleftrightarrow \mbox{(momentum space)} \Longleftrightarrow \nn\\
\delta\tilde{{\mathcal M}}(k,k^\prime) &=&
\iint \frac{d^4\hat k}{(2\pi)^4} \frac{d^4\hat k^\prime}{(2\pi)^4}~ 
\delta\tilde{{\mathcal D}}(\hat k,\hat k^\prime)~\tilde{{\bf K}}(\hat k^\prime,k^\prime,\hat k,k) 
\Longleftrightarrow \mbox{($\delta\tilde\chi$ is the gauge shift)} \Longleftrightarrow \nn\\ 
&i~e& \int d^4q/(2\pi)^4~\delta\tilde\chi(q)~
\left[\tilde{\mathcal M}(k+q,k^\prime)-\tilde{\mathcal M}(k,k^\prime-q)\right] ~=~ i~e 
\int d^4q/(2\pi)^4~\delta\tilde\chi(q)~ \times \nn\\ 
&\times& \iint \frac{d^4\hat k}{(2\pi)^4} \frac{d^4\hat k^\prime}{(2\pi)^4}~
\left[\tilde{\mathcal D}(\hat k+q,\hat k^\prime)-\tilde{\mathcal D}(\hat k,\hat k^\prime-q)\right]
~\tilde{{\bf K}}(\hat k^\prime,k^\prime,\hat k,k) 
\Longleftrightarrow \mbox{(homogeneous space-time$^{\ref{foot:DSE4_homo}}$)} \Longleftrightarrow \nn\\
&i~e~& \delta\tilde\chi(k^\prime-k) 
\left[\tilde{\mathcal M}(k^\prime)-\tilde{\mathcal M}(k)\right] ~=~ i~e~ 
\delta\tilde\chi(k-k^\prime) \int \frac{d^4\hat k}{(2\pi)^4}~
\left[\tilde{\mathcal D}(k + \hat k - k^\prime)-\tilde{\mathcal D}(\hat k)\right] \times \nn\\
&\times&\tilde{{\bf K}}(k + \hat k - k^\prime,k^\prime,\hat k,k) \;\; ;
\eea
after a variable shift ($k^\prime\to p+q,~ k\to p,~ \hat k\to k$), and accounting for the 
$\delta\tilde\chi$-oddness\footnote{The phase (gauge) shift oddness stems 
from the Green's function definition~:
${\displaystyle {\mathcal D}(x^\prime,x)~=~-{\mathcal D}(x,x^\prime)^\star}$.}, the proof is completed. 

The S-matrix unitarity naturally leads to physical conditions for the propagator 
structure, {\it i.e.} a pole term {\it times} a multiplicative constant 
(renormalization residue) \cite{BerestetskiiQED}. Within this context, although 
not needed\footnote{In principle, the set of equations (\ref{eqn:DSE4_homo}) accounts 
for an even number of unknowns ($\tilde{\mathcal M}$, $\tilde{\mathcal D}$, 
$\tilde\Pi$ and $\tilde{\bf K}$): it can be numerically solved self-consistently.}, 
we shall introduce the following ansatz on the mass operator:
${\displaystyle \tilde{\mathcal M}(p^2)=-\alpha \tilde D^{-1}(p^2) + \mu^2}$, being $\alpha$ 
a (real) constant to be determined, while $\mu$ is a mass scale (renormalization point). 
By this choice, the interacting propagator reads 
${\displaystyle \tilde{\mathcal D}^{-1}(p^2)=(1+\alpha)~\tilde D^{-1}(p^2) - \mu^2=
\frac{1+\alpha}{\alpha}\left[\frac{\mu^2}{1+\alpha}-\tilde{\mathcal M}(p^2)\right]}$, and the pole is 
located at ${\displaystyle \tilde D^{-1}\left( p^2_{pole} \right)\equiv\frac{\mu^2}{1+\alpha}}$. 
Moreover, by using this S-matrix unitarity inspired ansatz, the Bethe--Salpeter kernel 
$\tilde{\bf K}$ is promptly evaluated by means of the last equation in (\ref{eqn:DSE4_homo}):
\be
\label{eqn:BSK_eq}
\tilde{\bf K}(k-q,p+q,k,p) = \frac{\alpha}{1+\alpha}~\tilde{\mathcal D}^{-1}(k-q)~
\tilde{\mathcal D}^{-1}(k)~(2\pi)^4\delta^4(p-k+q)\;\ ,
\ee
by which the vertex $\tilde\Pi$ is shortened to ${\displaystyle (1+\alpha)~{\bf I}}$ 
(vertex renormalization) by direct substitution in the third equation of (\ref{eqn:DSE4_homo}).
 
Finally, the mass operator equation in (\ref{eqn:DSE4_homo}) accounts for the 
self-consistency of the Dyson--Schwinger approach:
\be
\label{eqn:DSE4_self-c}
\tilde{\mathcal M}(p^2) = (1+\alpha) \frac{\lambda^2}{2!}\int~d^4x~\exp\left(i~ p x\right) 
\left[{\mathcal D}(x)\right]^3 \;\ ,
\ee
where ${\mathcal D}(x)$ is the interacting propagator in space-time (configuration space), while 
the implicit dependence upon $\alpha$ in both $\tilde{\mathcal M}, {\mathcal D}$ is 
understood\footnote{
In euclidean four-dimensional space the interacting propagator, together with the mass operator 
ansatz, reads ${\displaystyle {\mathcal D}(x,m)=\frac{m x}{1+\alpha}\frac{K_1(m x)}{(2\pi)^2 x^2}}$, 
${\displaystyle m=\frac{\mu}{\sqrt{1+\alpha}}}$, ${\displaystyle x = \sqrt{x_\mu x^\mu}}$, 
being $K_1$ the first order modified Bessel 
function of second kind, while ${\displaystyle \tilde{\mathcal M}(p^2)=\alpha p^2 + \mu^2}$.
}. 
Eq.(\ref{eqn:DSE4_self-c}) describes a {\it setting sun} Feynman diagram: it provides either 
the broken symmetry solution or the symmetric one. To accomplish with the task, the mass 
operator of Eq.~(\ref{eqn:DSE4_self-c}) satisfies a dispersion relation (DR) with one 
subtraction \cite{GellMann:1954db,Wong:1957,Khuri:1957} 
(in $p$ variable)\footnote{Otherwise 
${\displaystyle \tilde{\mathcal M}_{sub}(p^2)=\tilde{\mathcal M}(p^2)-\tilde{\mathcal M}(0)}$, 
being ${\displaystyle \tilde{\mathcal M}(p^2)=\int_0^{+\infty}\frac{dt}{\pi}
\frac{{\tt Im}\left(\tilde M(t)\right)}{t-p^2}}$ the mass unsubtracted DR.}:
${\displaystyle \tilde{\mathcal M}_{sub}(p^2)=p^2 \int_0^{+\infty}\frac{dt}{\pi}
\frac{{\tt Im}\left(\tilde M(t)\right)}{t (t-p^2)}}$. 
We evaluate the subtracted version of Eq.~(\ref{eqn:DSE4_self-c}) on the mass pole, 
resembling the mass/gap equation of \cite{Nambu:PR122,Rajagopal:2000wf}. 
In euclidean space it reads \cite{Groote:1998wy}:
\be
\label{eqn:sunset}
m^2 = m^2~\frac{\lambda_R^2}{2! (2\pi)^4} \int_0^{+\infty}dx~
m~\left[\frac{1}{2}-\frac{J_1(m x)}{m x}\right]~K_1(m x)^3\times {\mathcal F}(x,\Lambda) \;\,
\ee
where $J_1$ is the first order Bessel function of first kind, 
${\displaystyle \lambda_R=\lambda/(1+\alpha)}$ and ${\displaystyle {\mathcal F}(x,\Lambda)}$ 
is an ultraviolet cut-off factor\cite{Nambu:PR122}. 
The integral of Eq.~(\ref{eqn:sunset}) has to be regularized at some point ($1/\Lambda$) since 
it exhibits a (logaritmic) divergence for small $x$ ($\Lambda\to\infty$) (the variable $x$ 
behaves like the inverse four-momentum $p$, thus at the origin $x \sim 1/\Lambda$). 
 
After rescaling the variable $x\to \xi/m$ and normalizing 
the parameter $m$ to the cut-off $\Lambda$ ($m\to\eta~\Lambda$), Eq.~(\ref{eqn:sunset}) reduces 
to\footnote{Hereafter we implement 
${\displaystyle {\mathcal F}(x,\Lambda)=\theta\left[x-1/\Lambda\right]}$, being $\theta$ the Heaviside 
function.}:
\be
\label{eqn:fixedpoint}
\eta^2 = \eta^2~\frac{\lambda_R^2}{2! (2\pi)^4} \int_\eta^{+\infty}d\xi~
\left[\frac{1}{2}-\frac{J_1(\xi)}{\xi}\right]~K_1(\xi)^3 \;\ .
\ee
For vanishing $\eta$ the mild divergence of the integral on the r.h.s. of Eq.~(\ref{eqn:fixedpoint}) 
is compensated by the $\eta^2$ term and the trivial solution ($\eta=0$) is achieved.
 
The importance of Eq.~(\ref{eqn:fixedpoint}) 
is twofold. On one hand, a mapping between $\lambda_R$ and $\eta$ can be obtained beyond the 
perturbative regime of the coupling constant, helpful while investigating the cohexistence of 
the triviality (namely $\lambda_R\to 0$ when $\Lambda\to\infty$) 
and the spontaneous symmetry breaking (SSB) of the theory \cite{Ladisa_preprint} (crucial, 
for instance, for the self-consistency of the Standard Model in particle 
physics \cite{DjouadiPR457}).
 
On the other hand, a renormalization group equation (RGE) for the $\eta$-evolution of the coupling 
constant $\lambda_R$ can be promptly assessed from Eq.~(\ref{eqn:fixedpoint}):
\be
\label{eqn:RGE}
\lambda_R^2(\eta)=
\frac{\lambda_R^2(\eta_0)}
{1+\frac{\lambda_R^2(\eta_0)}{2! (2\pi)^4} \int_{\eta}^{\eta_0}dx~\Omega(x)} 
\stackrel{\eta\to 0,~\eta_0/\eta~ fixed}{\approx}
\frac{\lambda_R^2(\eta_0)}
{1+\frac{\lambda_R^2(\eta_0)}{2! (4\pi)^4} \ln\left(\frac{\eta_0}{\eta}\right)} \;\ ,
\ee
being $\Omega(x)$ the integrand function of Eq.~(\ref{eqn:fixedpoint}).
While the latter equation corresponds to the perturbative region for the 
coupling $\lambda_R$, the former one holds in non-perturbative regime too. 
 
Beyond the trivial solution ($\eta=0$), corresponding to the unbroken symmetry \cite{CooperPRD70}, 
Eq.~(\ref{eqn:fixedpoint}) provides with the broken symmetry solution, once the renormalized 
coupling $\lambda_R$ is replaced according to the SSB mechanism for the theory investigated. 
Indeed both $m^2$ and $\lambda_R$ depend upon the renormalization constant $\alpha$ in the 
same fashion ${\displaystyle \left(\propto \frac{1}{1+\alpha}\right)}$; 
thus, by eliminating it, we get the formula 
${\displaystyle m^2 = \frac{\mu^2}{\lambda}~ \lambda_R}$, accounting for the not vanishing vacuum 
expectation value of the field which minimizes the Hamiltonian 
${\displaystyle \left(\langle 0|\phi^\star \phi|0\rangle \stackrel{def.}{=} \upsilon^2 \not = 0, 
\frac{\mu^2}{\lambda} = \frac{\upsilon^2}{2}\right)}$.
It reads a typical fixed point problem for the parameter $\hat m$ 
${\displaystyle \left(\hat m \stackrel{def.}{=} \frac{m}{\upsilon},~ 
\hat \Lambda \stackrel{def.}{=} \frac{\Lambda}{\upsilon}\right)}$ 
\cite{BanachFM3,CaccioppoliRANL11}:
\be
\label{eqn:fixedpoint2}
\hat m = 4 \pi \sqrt[4]{ \frac{\rho}{2 \ln(\hat m)}}\;\; ,
\ee 
being ${\displaystyle \rho \stackrel{def.}{=} 
\left[\frac{\lambda_R(\hat m/\hat \Lambda)}{\lambda_R(1/\hat\Lambda)}\right]^2}$. 
Here the parameter $\rho$ can be computed by means of the Eq.~(\ref{eqn:fixedpoint}). 
In the limit $\hat\Lambda\to\infty$ it approaches the unity regardless of the 
actual value for $\hat m$: for instance, we find $\rho \in [1.019,1.072]$ for 
$\hat m \in [2,12]$ (at $\hat\Lambda \simeq 4\times 10^{18}$).
 
Implementing the Eq.~(\ref{eqn:fixedpoint2}) is straighforward 
and computationally not demanding. 
In spite of its simplicity, it clearly exhibits the iterative solution scheme, following the 
spirit of the typical condensed matter approach to the solution of the Dyson--Schwinger equations 
original sets (\ref{fig:DSE4},\ref{eqn:DSE4_homo}). 
 
We find $\hat m = 8.88\pm 0.10$, where the error comes from the parameter $\rho$ ranging in 
the aforementioned interval. Our finding is in agreement with the value of 
$\hat m = 2\sqrt{2}\pi\simeq 8.88577$ predicted by the classically scale invariant 
(CSI) theory and computed 
on the lattice \cite{Consoli:1993jr}. Indeed, in CSI theory an effective potential 
is computed starting from the euclidean action of a massless self-interacting 
scalar field and integrating out the field fluctuations around its 
vacuum expectation value (VEV). Different schemes are possible: 
in Ref.~\cite{Consoli:1993jr}, for instance, two different renormalization constants 
are introduced for the field VEV and for the fluctuations. Another possibility 
is to add a mass term in the euclidean action and to describe the 
broken-to-symmetric phase as a phase transition. In order to account for different 
phenomenological approaches, the effective potential depends on a parameter 
ranging between 1 (CSI case) and 3 (classical quartic potential), while the value 
2 is a phase transition signature. The effective potential is then expanded up to 
the third power around its minimum and the three coefficients of the expansion, 
related to the abovementioned parameter, are fitted by the lattice simulation data. 
The lattice data fitted central value is 1.14, although the resulting uncertainty 
is large (see Ref.~\cite{Consoli:1993jr} and references therein for details). 
While the latter statement signals the need for more statistics in lattice 
calculations, the former result points at the CSI theory as the preferred scenario.

\par
In conclusion, we believe that the Dyson--Schwinger 
equations in QED, together with the Green--Ward--Takahashi identity, 
we investigate in this paper, are equivalent to the analogous 
set of integral equations studied in condensed matter, often 
referred to as the Hedin's equations. They account for the 
self-consistency of the method, corresponding to the 
truncation of the perturbative/diagrammatic expansion in 
particle physics. 
Within the scheme proposed, the approach seems to be 
capable of producing the non-perturbative solution of the 
self-interacting scalar field theory, as suggested in the CSI theory, 
within theoretical errors. 
This method could be straighforwardly applied to other abelian 
(as well as non-abelian) theories, for instance the self-interacting 
fermion field theory, as an effective theory for describing the gap 
opening (mass generation) mechanism in condensed matter. 
We only mention here a result for 
the graphene, a system composed by sp$_2$-hybridized carbon atoms 
placed on a plane. Such a system is interesting both on experimental 
and theoretical side, due to its peculiar electronic 
properties (see, for instance, Ref.~\cite{Trevisanutto:2008prl} and references therein). 
While from the former point of view the gap opening is still 
under investigation, it is not theoretically understood yet whether 
or not its spatial dimensionality is two (as assumed in all 
the theoretical works) or three (as suggested by the role 
played by the off plane p$_z$ orbitals). Both issues (gap and 
dimensions) could be addressed by a Dyson--Schwinger approach together 
with a similar mass operator ansatz: indeed the mass/gap 
equation, as it appears in the Eq.~(\ref{eqn:DSE4_self-c}), 
actually depends on the space-time dimension \cite{Groote:1998wy}. 
Therefore, if a gap opens, it will depend on the actual spatial 
dimension (probably between two and three). The details of this work 
will be given elsewhere together with the results.
%
% 
%{\it Acknowledgements} xxx thank(s) yyy for useful correspondence 
%(collaboration) on the zzz (in the early stage of this work).

\end{document}